\documentclass[twocolumn,prl,showpacs,amsmath,amssymb]{revtex4}

\usepackage{graphicx}
\usepackage{dcolumn}
\usepackage{bm,color}

\preprint{APS/}
\begin{document}

\title{Synthesis of thin-long heavy nuclei in ternary collisions}

\author{Yoritaka Iwata$^{1}$}
\author{Kei Iida$^{2}$}
\author{Naoyuki Itagaki$^{3}$}
 \affiliation{$^{1}$GSI Helmholtzzentrum f\"ur Schwerionenforschung, D-64291 Darmstadt, Germany \\
 $^{2}$Department of Natural Science, Kochi University, Akebono-cho, Kochi 780-8520, Japan \\
 $^{3}$Yukawa Institute for Theoretical Physics, Kyoto University, Kyoto 606-8502, Japan}

\date{\today}

\begin{abstract}
We illustrate the formation of a thin-long structure of heavy nuclei 
by three-nucleus simultaneous collisions within time-dependent density functional theory.  
The impact parameter dependence for such 
formation is systematically demonstrated through clarifications of the difference 
between binary and ternary collision events.  
A new method for producing thin-long 
heavy nuclei in the laboratory is suggested, as well as the possible formation of the thin-long structure in hot dense matter such as that encountered in core 
collapse supernovae.  
\end{abstract}


\maketitle


Generally, synthesis of superheavy elements and many other exotic nuclei 
comes from collisions between two stable nuclei, whereas simultaneous 
collisions among more than two stable nuclei, which can occur in principle, 
have not been taken seriously.  In experiments using accelerators, it seems 
difficult to construct a setup to make three nuclei collide 
simultaneously.  Nevertheless it is reasonable to expect that a fixed target 
can be bombarded by two beams moving in the opposite directions.  
In astrophysical circumstances, on the other hand, there seem many chances to 
see three-body reactions such as triple-alpha reactions \cite{clayton}.  
The triple-alpha reactions, if occurring through two-body resonances as in
the Hoyle picture, are similar to radioactive ion beam experiments 
\cite{rnb7} in the sense that two reactions occur non-simultaneously in 
terms of a time scale of the strong interaction.  
Even in stars and supernova cores, however, simultaneous three-nucleus collisions 
within the strong interaction time scale are usually presumed to be rather hard 
to encounter, let alone more than three nucleus collisions.  
How hard they are remains to be examined thoroughly, although the possible influence 
of simultaneous triple-alpha fusion reactions on carbon production in 
stars has attracted much attention \cite{ogata}.

Two-nucleus fusion reactions in stars and supernova cores are often 
investigated in two steps \cite{ichimaru,sawyer}.  One starts with 
fundamental two-nucleus fusion reactions in vacuum and then incorporates 
medium (plasma) effects.  For example, in the famous Gamow approach, one 
solves the elementary tunneling problem for given relative kinetic energy 
and then obtain the temperature-dependent reaction rate by averaging 
the resulting kinetic-energy-dependent reaction rate over the Boltzmann 
distribution of the kinetic energy.  
In this work, we concentrate on the 
fundamental binary and ternary collisions within the 
time-dependent density functional theory (TDDFT).  
Among multi-nucleus collisions, ternary collisions 
are particularly worth investigating in the sense that there always 
exists one plane containing all the three center-of-mass coordinates 
of colliding pairs.  Such a geometric restriction facilitates 
the formation of a low-dimensional quantum system.
As we shall see, 
a thin-long structure of the fusion products can be stabilized by 
rotation in the case of non-central binary and ternary collisions.

The TDDFT approach, which was originally proposed by 
P.~A.~M.~Dirac~\cite{dirac}, is useful for describing nuclear collisions 
at low energies except the reactions below the Coulomb barrier.  In the 
present case, we perform three-dimensional TDDFT calculations with Skyrme-type 
effective nucleon-nucleon interactions (SLy6~\cite{Chabanat-Bonche} and 
SKI3~\cite{reinhard}).  One can then derive various quantities from the many-nucleon 
wave function self-consistently obtained in the form of the Slater determinant.  
We remark that the colliding system would break into a few 
fragments with various neutron/proton ratios~\cite{iwata-prl} for the 
collision energy per nucleon of order the nucleon Fermi energy, and 
multi-fragmentation takes place for rather higher energies.  Multi-fragmentation, 
in which two-nucleon collisions are usually expected to play an 
important role \cite{PTEP}, is out of our scope.

\begin{figure}
\includegraphics[width=6cm]{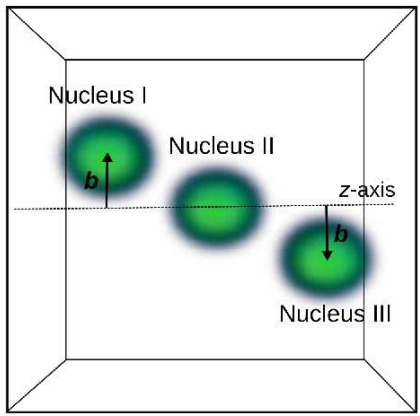} 
\caption{\label{fig2} (Color online)
The initial positions of three identical $^{56}$Fe nuclei, Nucleus I, 
Nucleus II, and Nucleus III, which are set to $(|{\bm b}|,0,15~{\rm fm})$, 
$(0,0,0)$, and $(-|{\bm b}|,0,-15~{\rm fm})$, respectively.
The velocity vector of each nucleus is given as 
$(0,0,|{\bm v}|)$, $(0,0,0)$, and $(0,0,-|{\bm v}|)$, respectively.
The volume of a box inside which the simulation is performed is 
$48 \times 48 \times 48$ fm$^3$.  
}
\end{figure}

\begin{figure} 
(a) $|{\bm b}| = 1$~fm  \\
\includegraphics[width=2.cm]{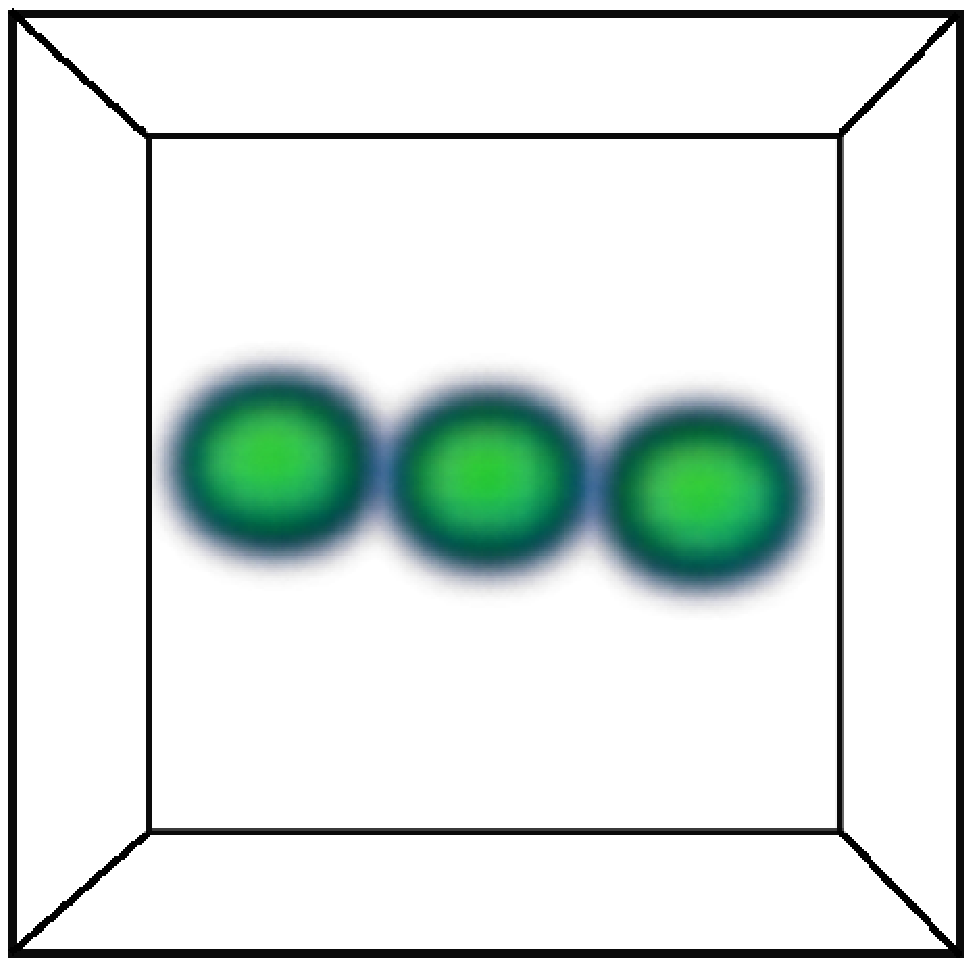}
\includegraphics[width=2.cm]{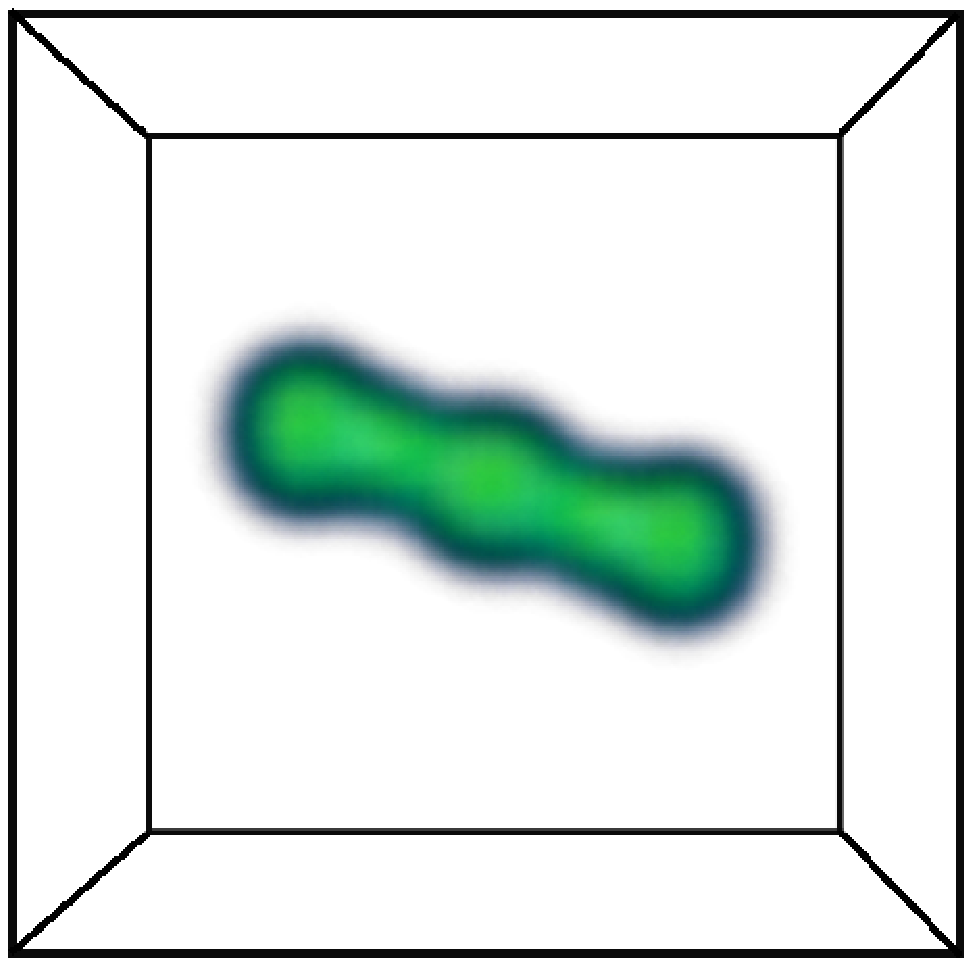}
\includegraphics[width=2.cm]{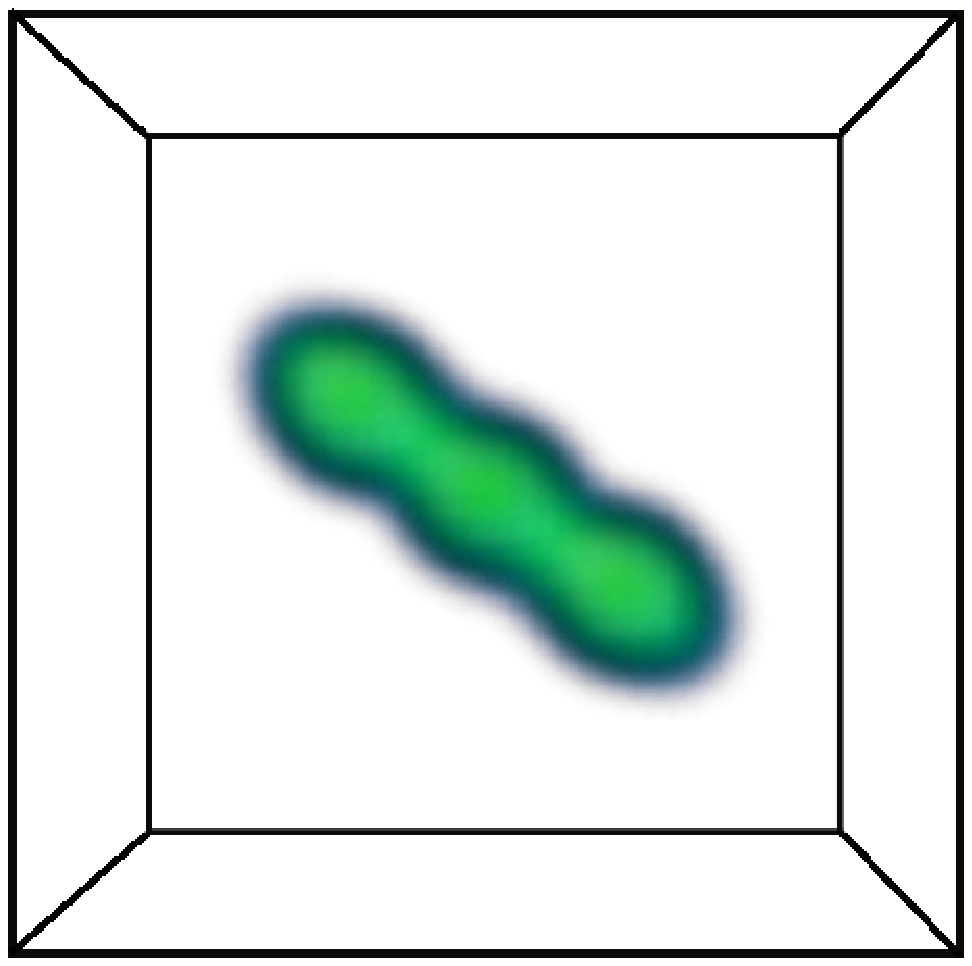}\vspace{1.5mm} \\ 
(b) $|{\bm b}| = 3$~fm  \\ 
\includegraphics[width=2.cm]{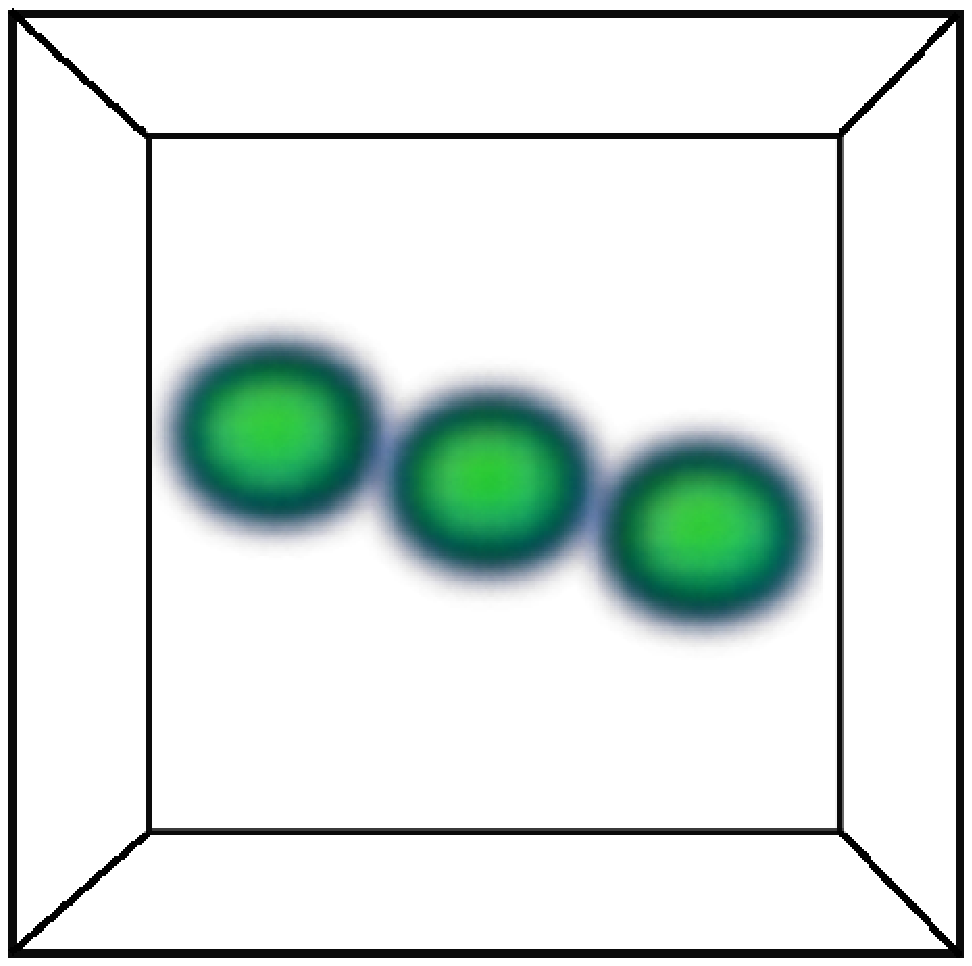}
\includegraphics[width=2.cm]{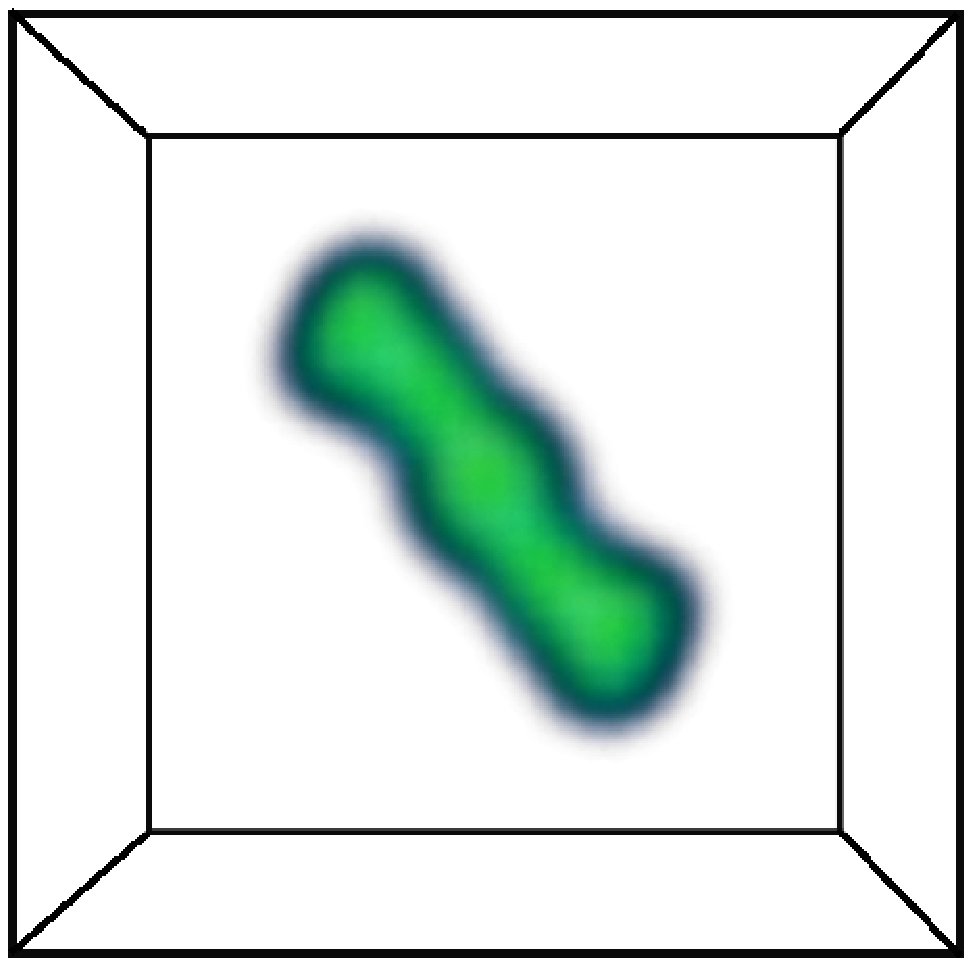}
\includegraphics[width=2.cm]{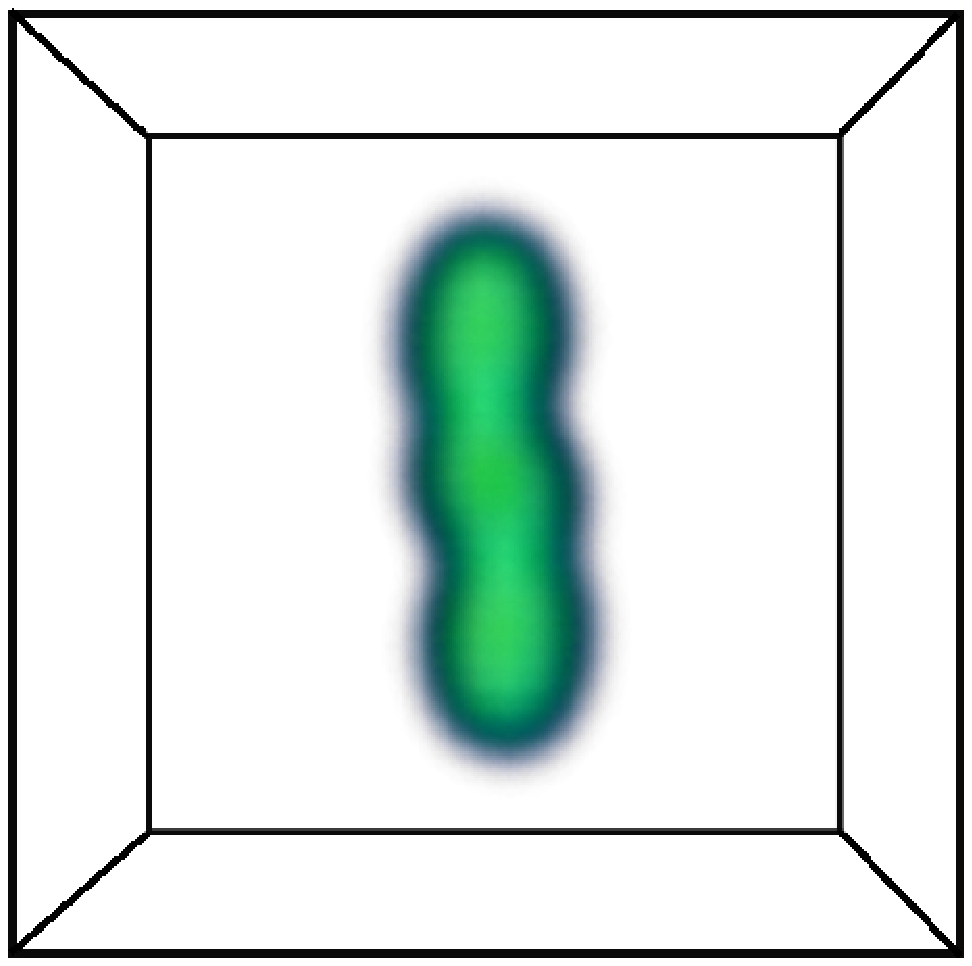}\vspace{1.5mm} \\ 
(c) $|{\bm b}| = 5$~fm  \\ 
\includegraphics[width=2.cm]{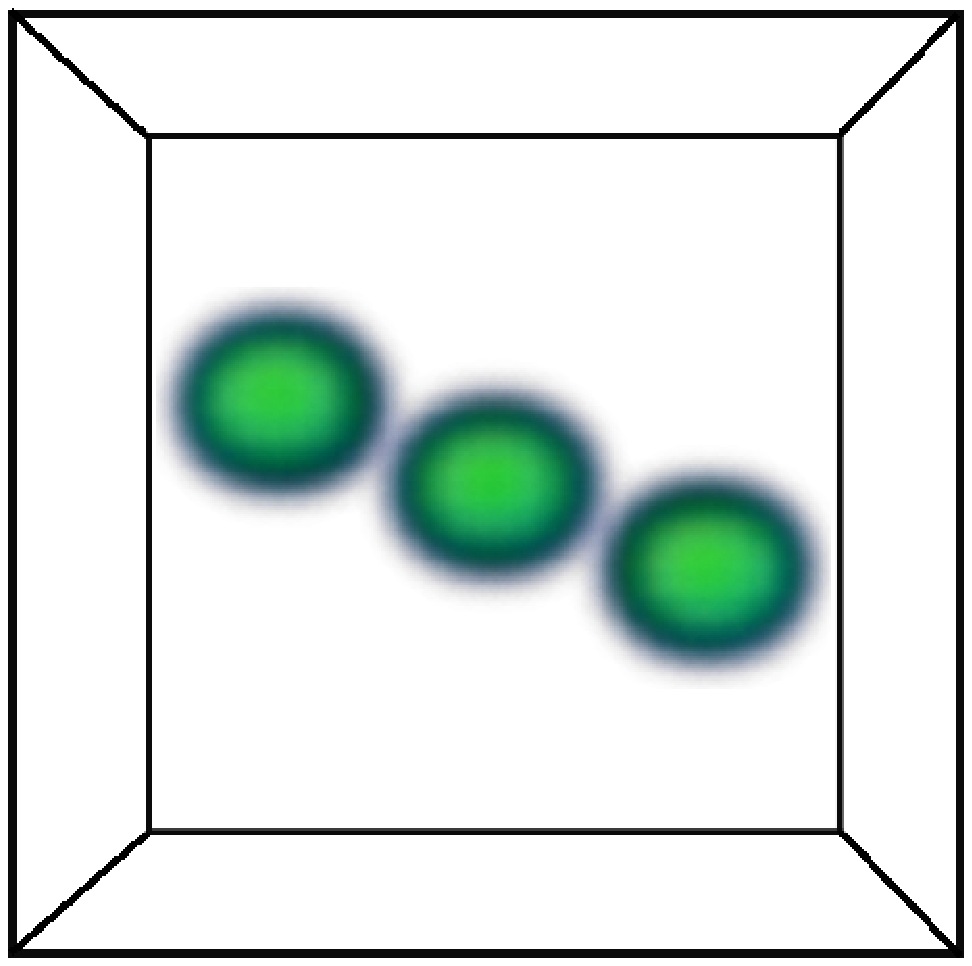}
\includegraphics[width=2.cm]{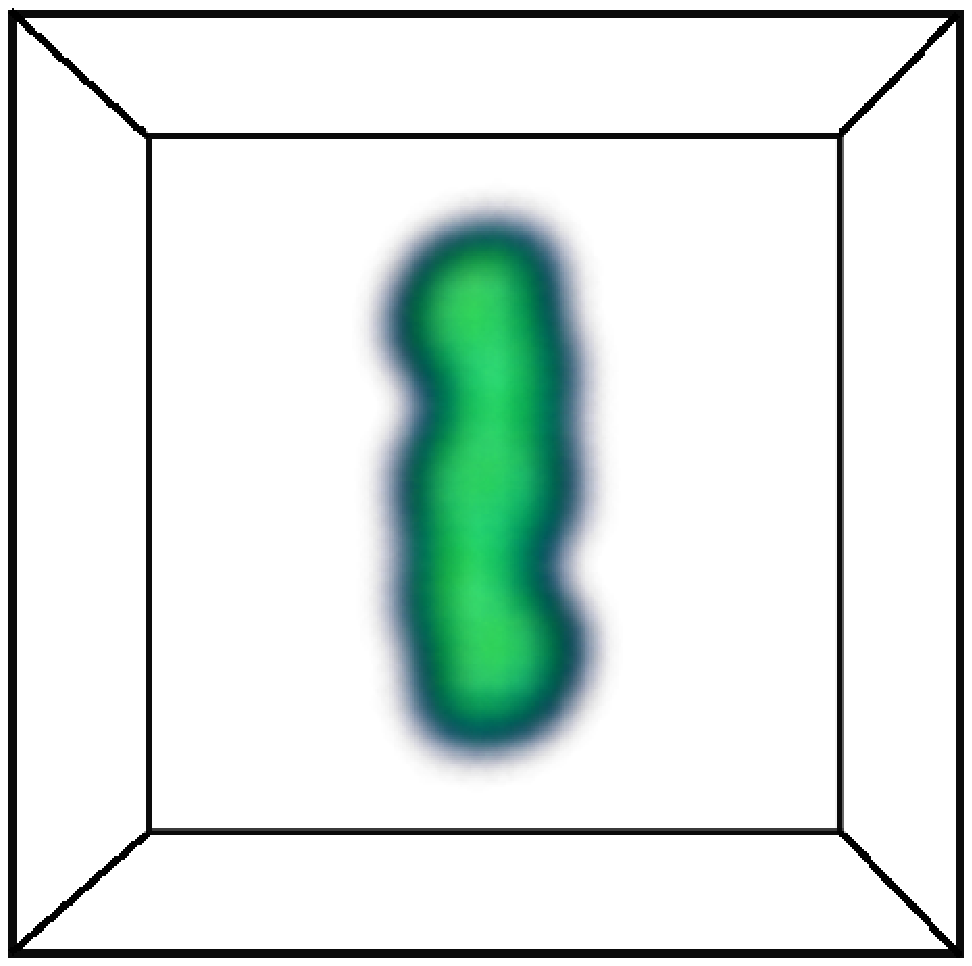}
\includegraphics[width=2.cm]{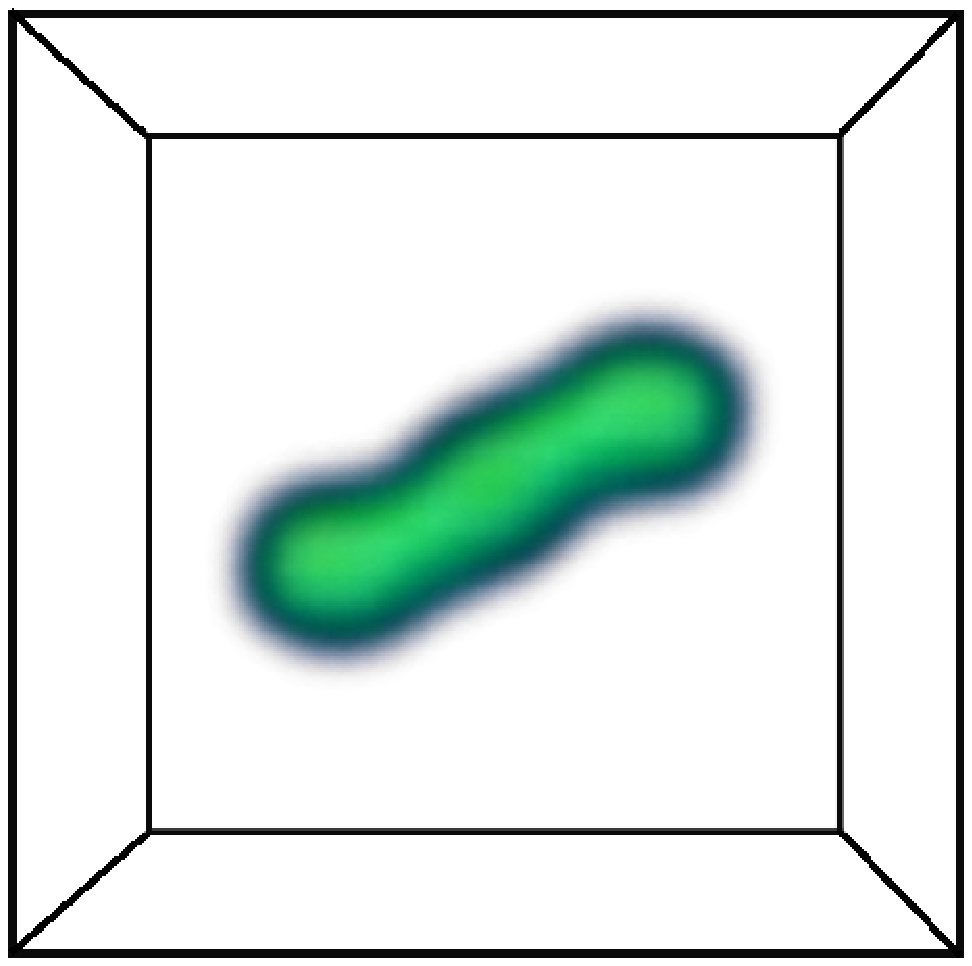}\vspace{1.5mm} \\ 
(d) $|{\bm b}| = 7$~fm  \\ 
\includegraphics[width=2.cm]{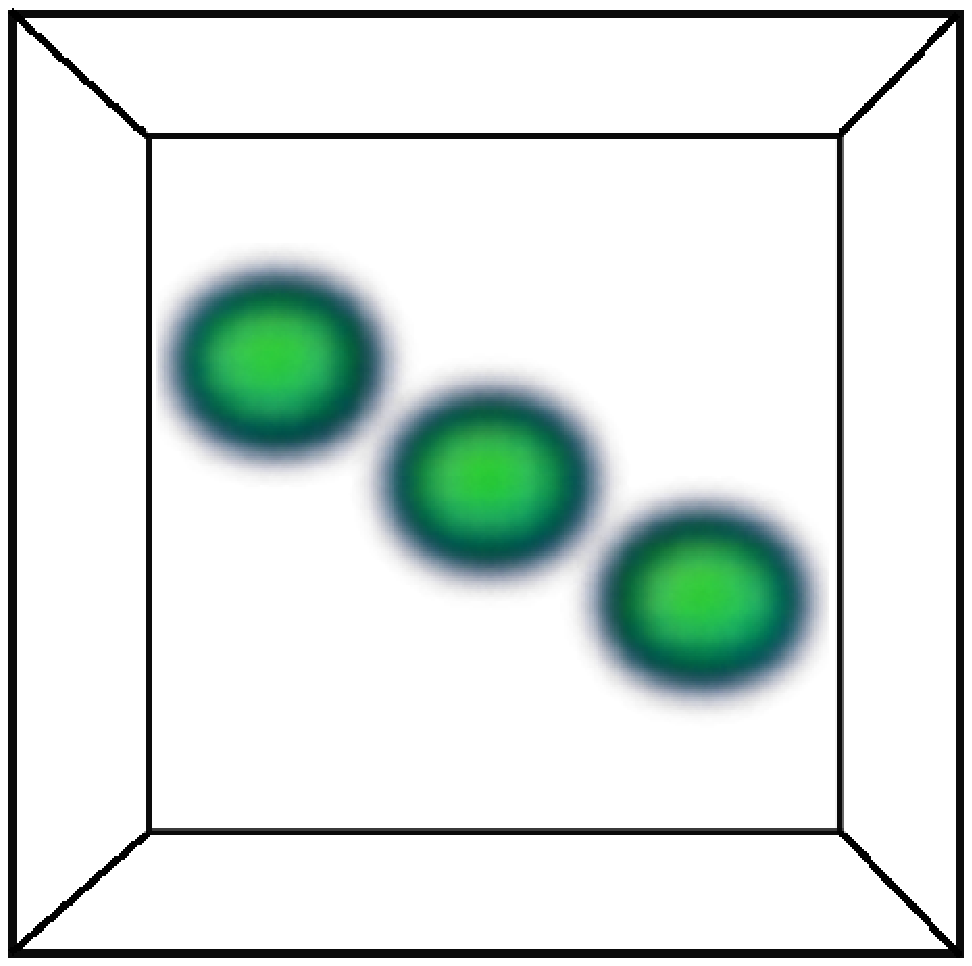}
\includegraphics[width=2.cm]{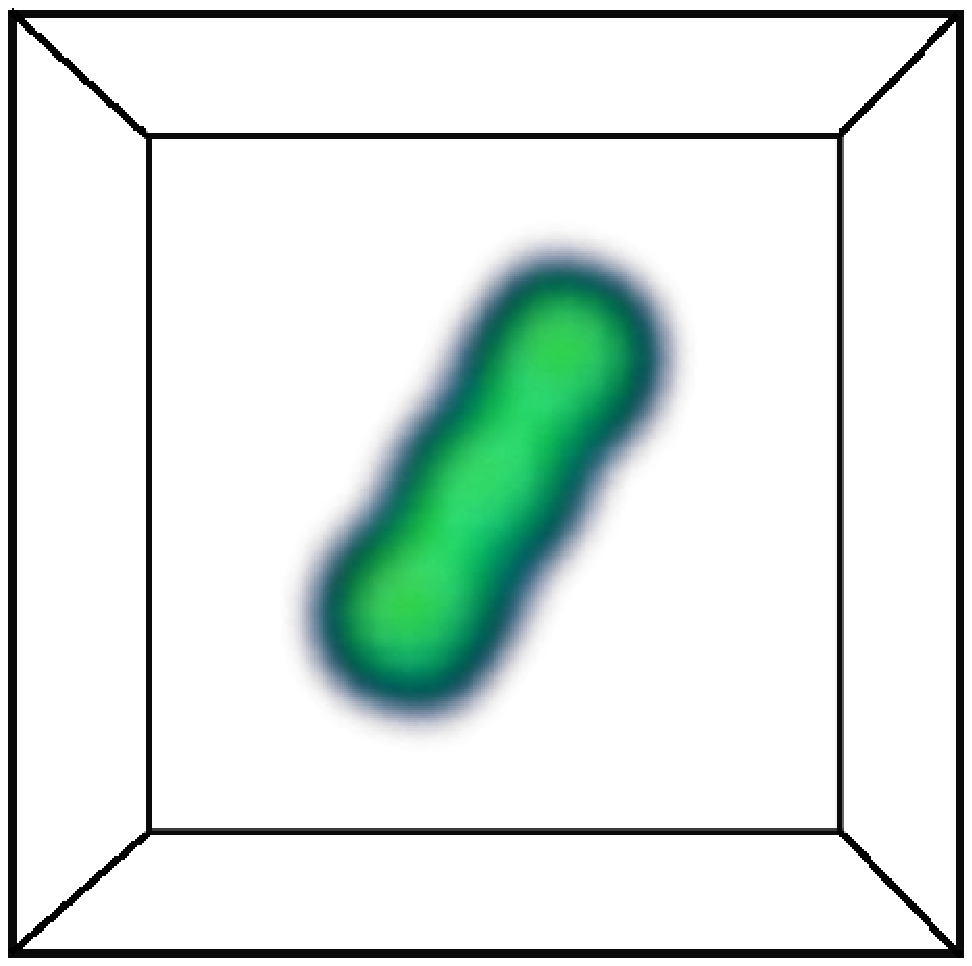}
\includegraphics[width=2.cm]{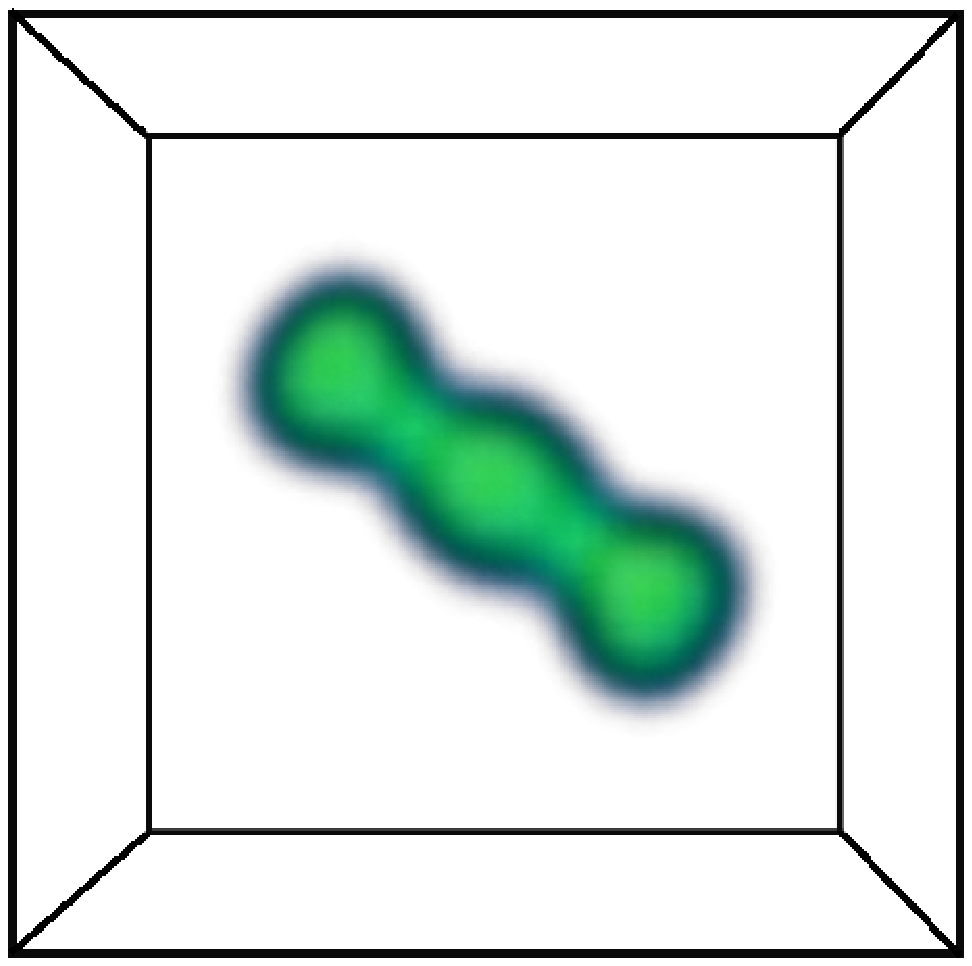}\vspace{1.5mm} \\
(e) $|{\bm b}| = 9$~fm  \\ 
\includegraphics[width=2.cm]{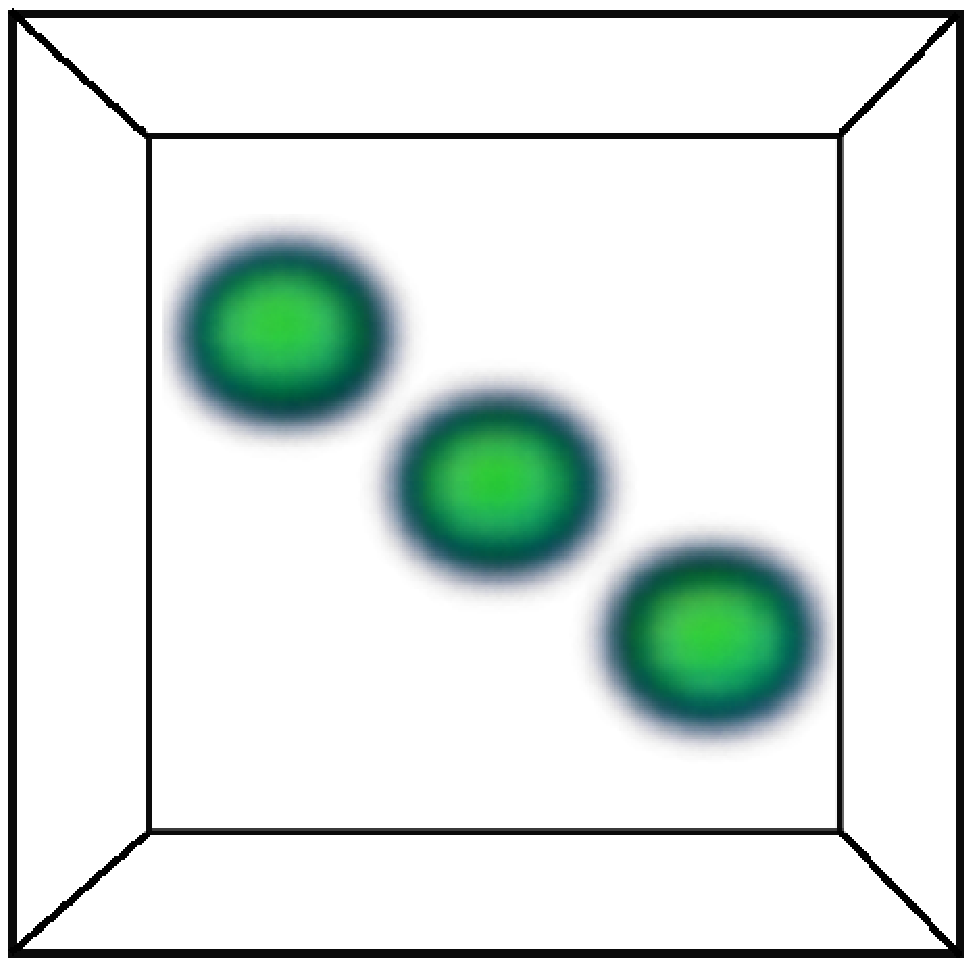}
\includegraphics[width=2.cm]{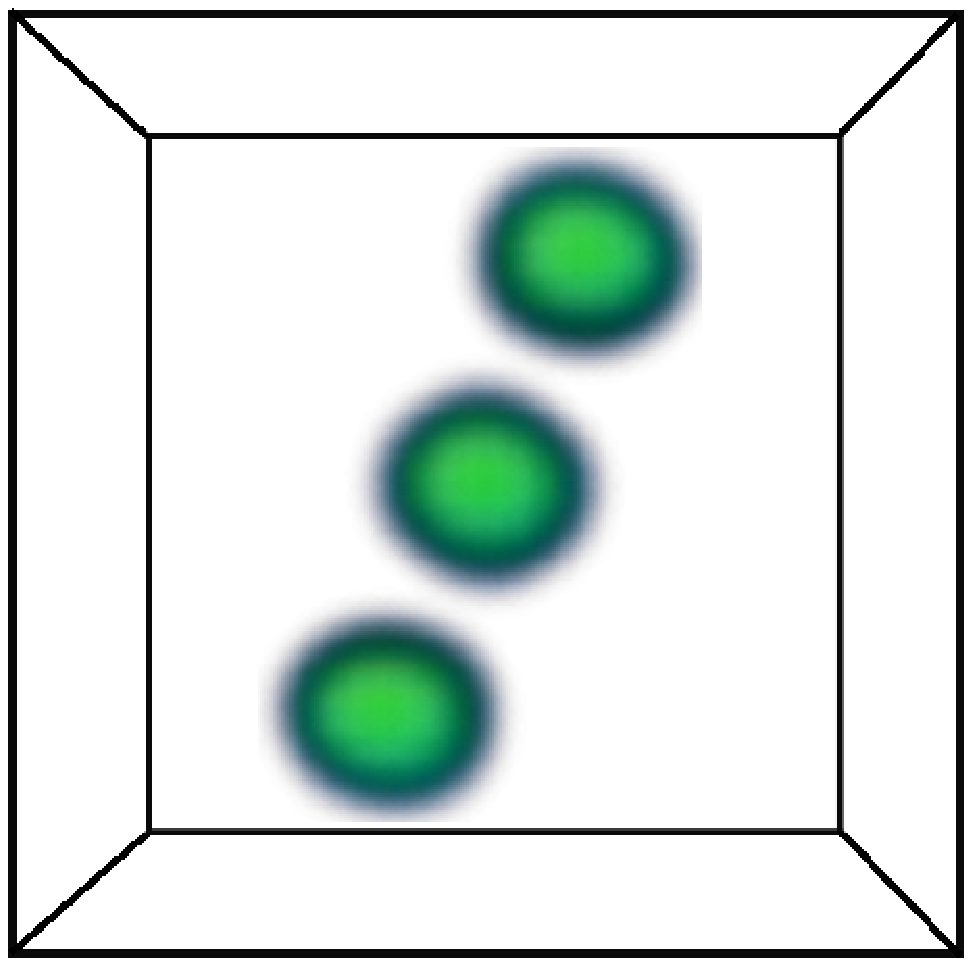}
\includegraphics[width=2.cm]{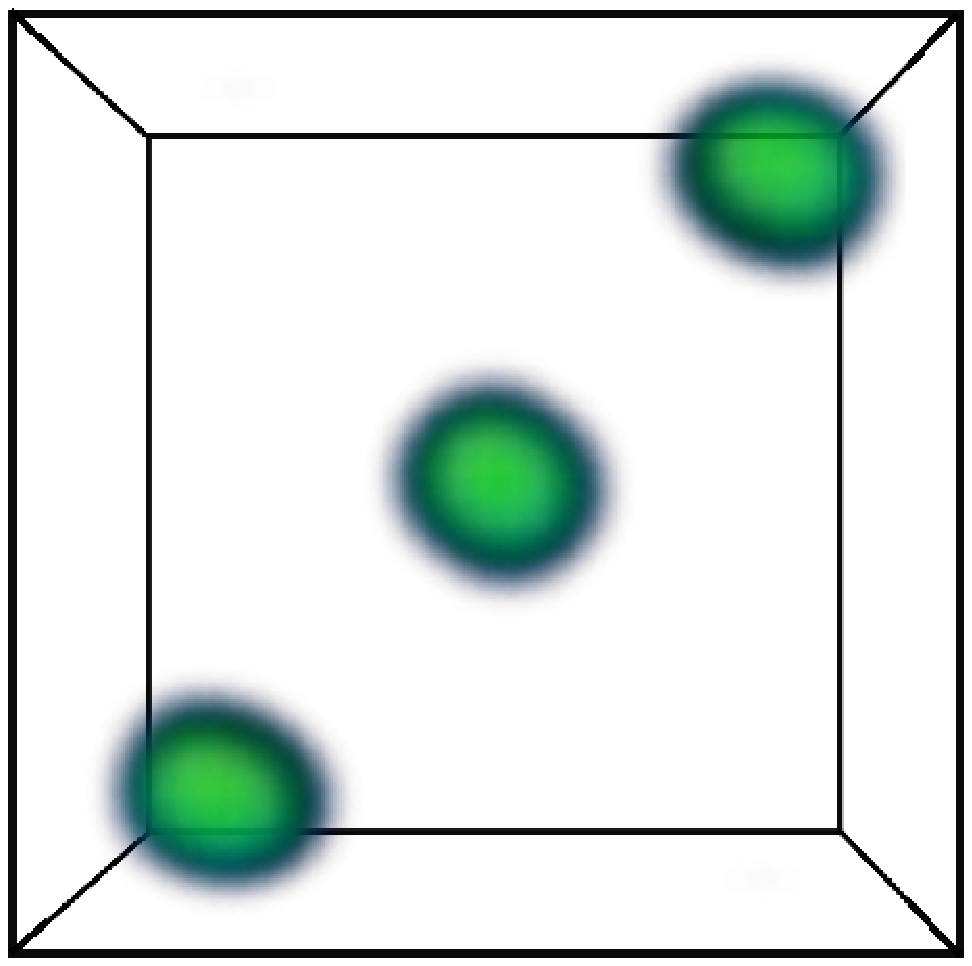}\vspace{1.5mm} \\ 
\caption{\label{fig3}(Color online)
Ternary collisions with different impact parameters 
that start with the initial condition shown in Fig.~\ref{fig2} 
(the SLy6 parameter set is taken).
The velocity is set to be $1.95 \times 10^{22}$~fm/s. 
Snapshots at 1.33$\times10^{-22}$~s (left), 13.3$\times10^{-22}$~s (middle), 
and 26.6$\times10^{-22}$~s (right) are shown for $|{\bm b}|$ = 1, 3, 5, 
7~fm, while those at 1.33$\times10^{-22}$~s, 13.3$\times10^{-22}$~s, 
and 20.0$\times10^{-22}$~s are shown for $|{\bm b}|$ = 9~fm.}
\end{figure}

Since $^{56}$Fe can be abundant in dense stellar matter, we consider 
heavy element synthesis due to multiple $^{56}$Fe collisions within $\sim 33.3 
\times 10^{-22}$~s, which corresponds to the typical duration 
time of low-energy heavy-ion reactions.  We first show an example of ternary collisions with the initial condition shown in 
Fig.\ \ref{fig2} and the incident energy of $1.47$~MeV per nucleon in 
the center-of-mass frame, which is illustrated in Fig.\ \ref{fig3}. 
In this case, the fusion product is $^{168}$Pt.
In contrast with reactions producing light elements, the corresponding 
reactions are endothermic and hence require additional energy even after 
the contact.  For the impact parameters that allow Nuclei I and II 
as well as Nuclei II and III to touch with each other, corresponding
to the cases of $|{\bm b}|=1,3,5,7$ fm, the final products have a thin-long 
structure of length 25-30 fm, which is stabilized by rotation.  
For example, in the case of $|{\bm b}| = 7$~fm, rotation up to $\pi$ 
rad is achieved in a few 10$^{-21}$~s, which corresponds to 
the total rotational energy of order $2.10$~MeV.  Note that 
thin-long nuclei can occur at optimal values of the incident energy, 
below and above which fusion is drastically suppressed.  Note also 
that the stabilization of such a thin-long structure by rotation 
can be seen in exotic clustering of light nuclei \cite{ichikawa11}.

It has been shown here that thin-long structures are produced 
by non-central ternary collisions for a broad range of the impact 
parameter and that the rotation speed becomes faster for larger 
$|{\bm b}|$ as it should.  Note that thin-long structures turn into 
more spherical shapes within $33.3 \times 10^{-22}$~s if there are no 
rotations, which suggests that the rotation plays an indispensable 
role in keeping the fusion products thin-long.

\begin{figure}
(a) $|{\bm b}| = 1$~fm  \\
\includegraphics[width=2.cm]{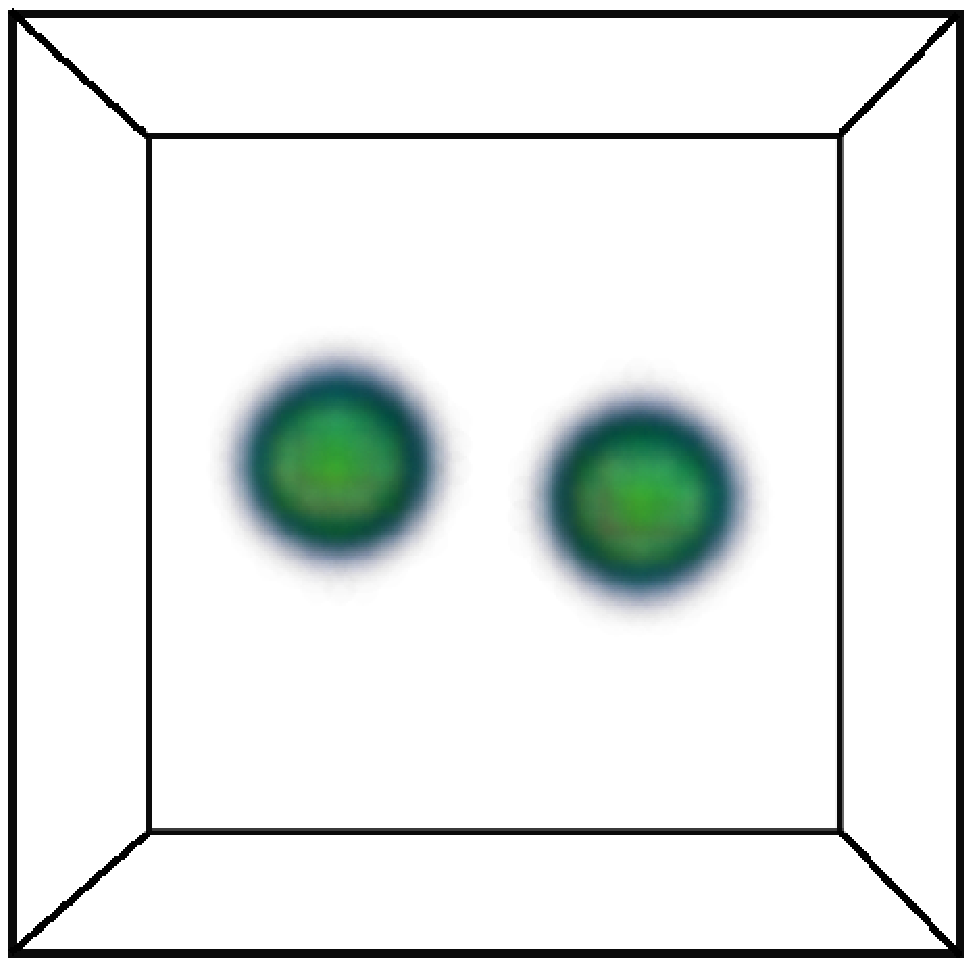}
\includegraphics[width=2.cm]{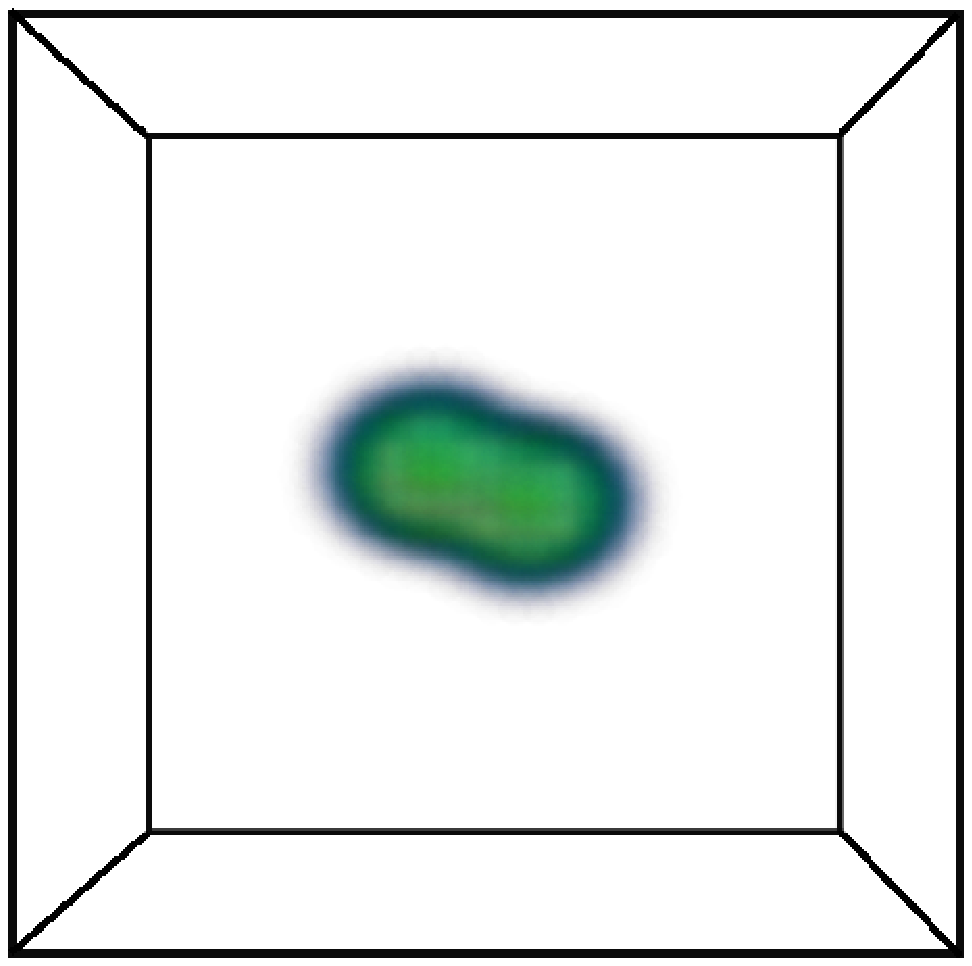}
\includegraphics[width=2.cm]{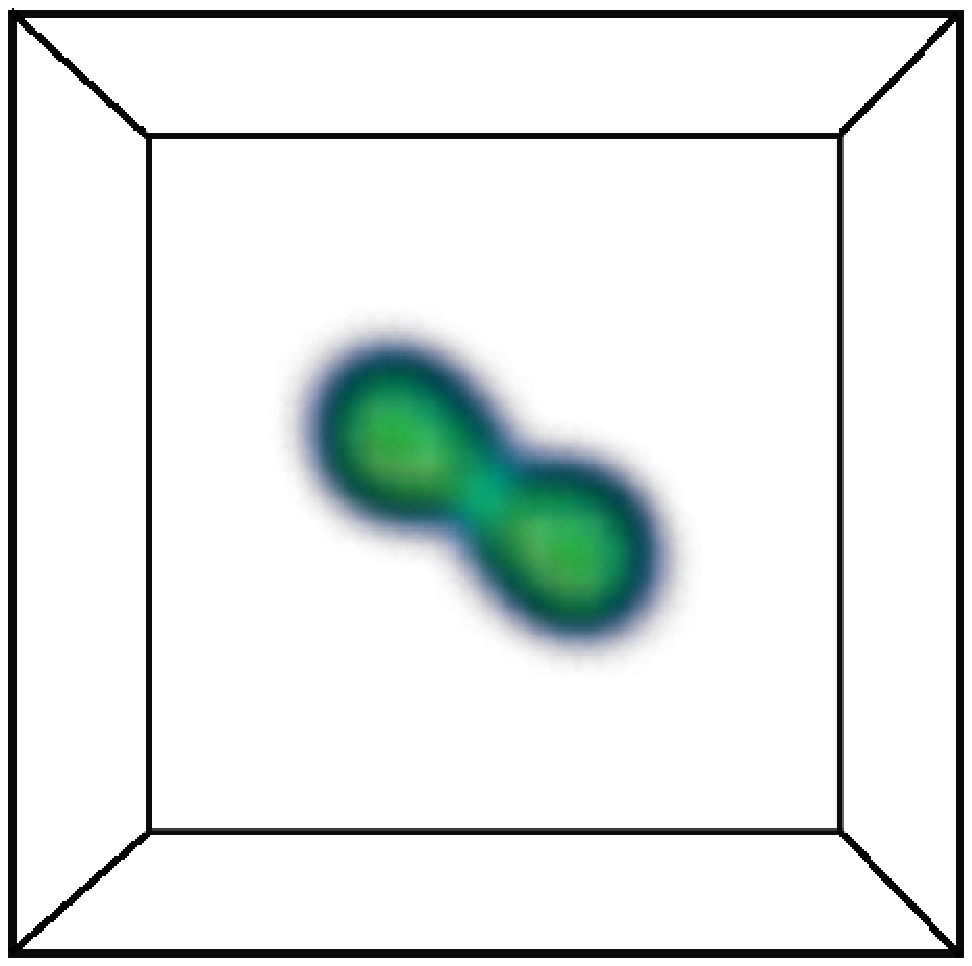}\vspace{1.5mm} \\ 
(b) $|{\bm b}| = 3$~fm  \\ 
\includegraphics[width=2.cm]{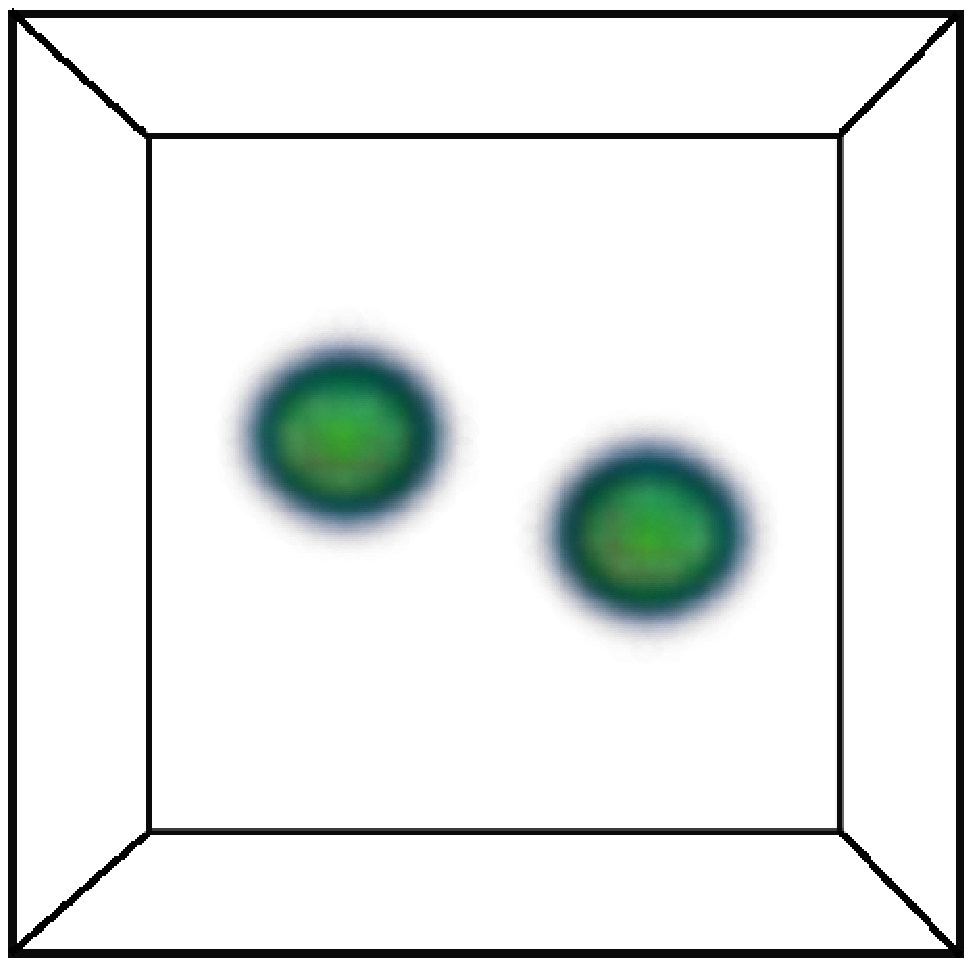}
\includegraphics[width=2.cm]{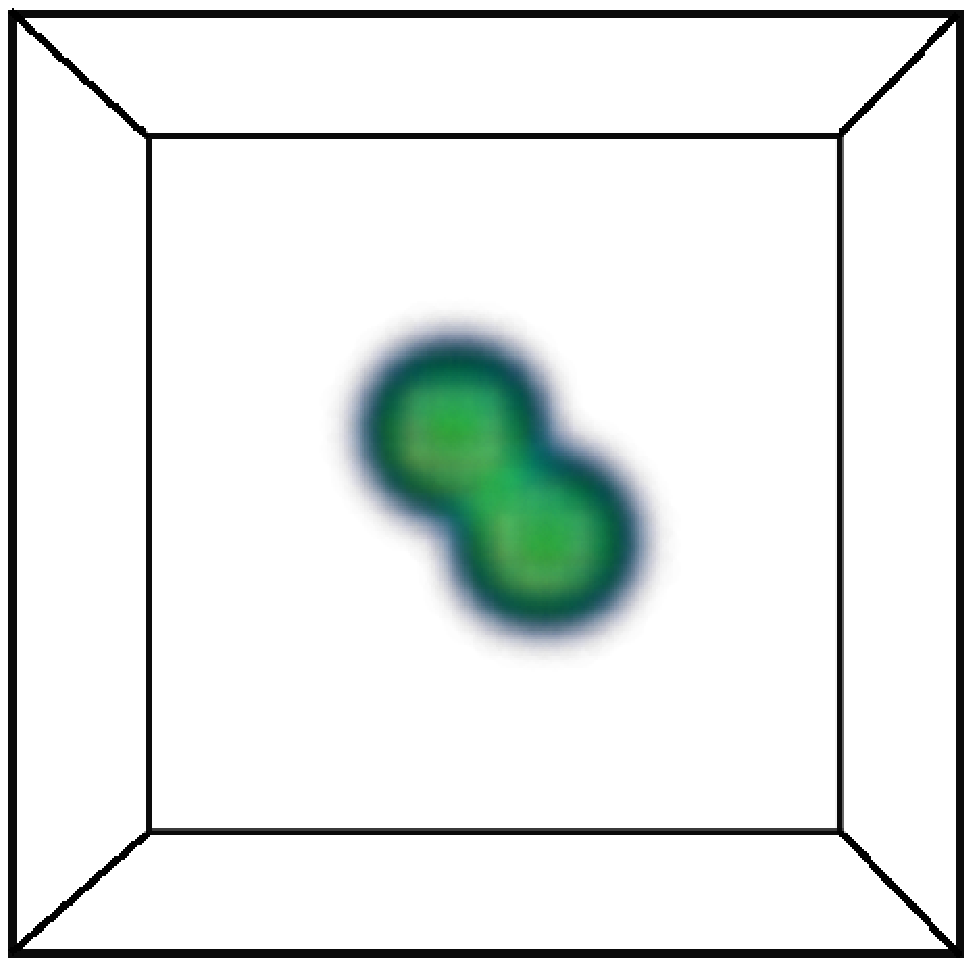}
\includegraphics[width=2.cm]{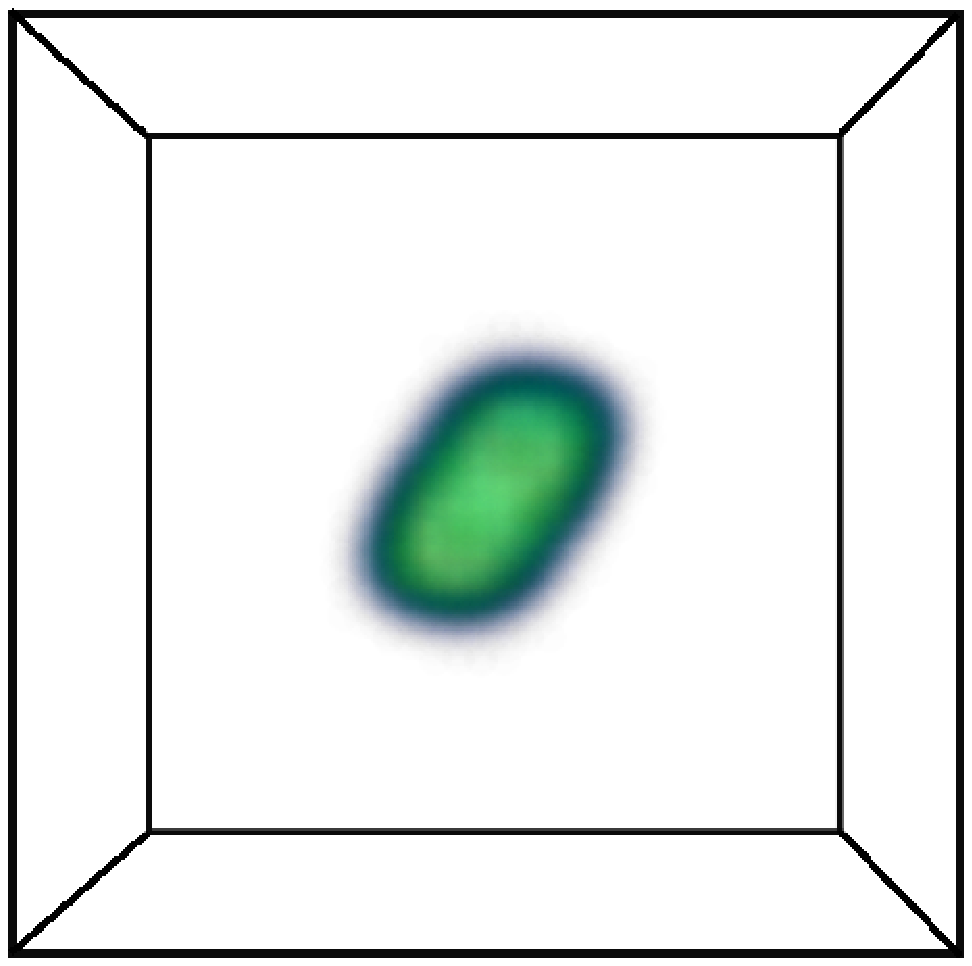}\vspace{1.5mm} \\ 
(c) $|{\bm b}| = 5$~fm  \\ 
\includegraphics[width=2.cm]{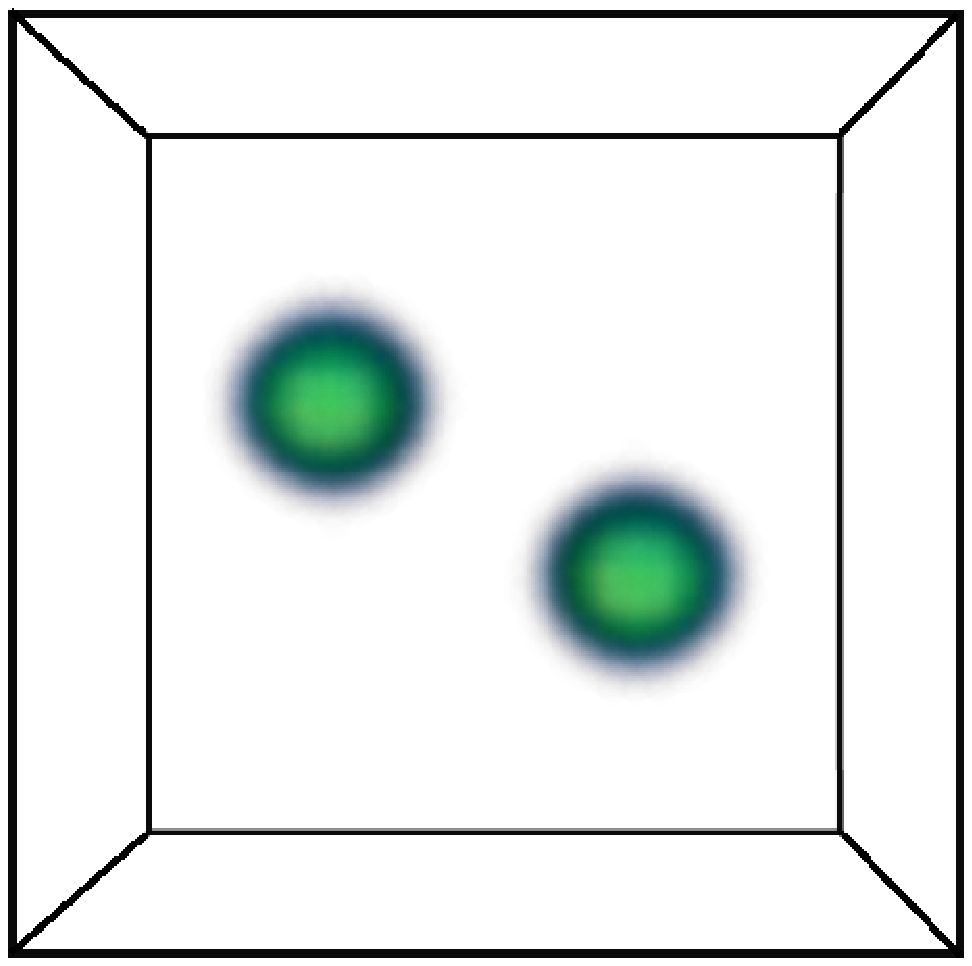}
\includegraphics[width=2.cm]{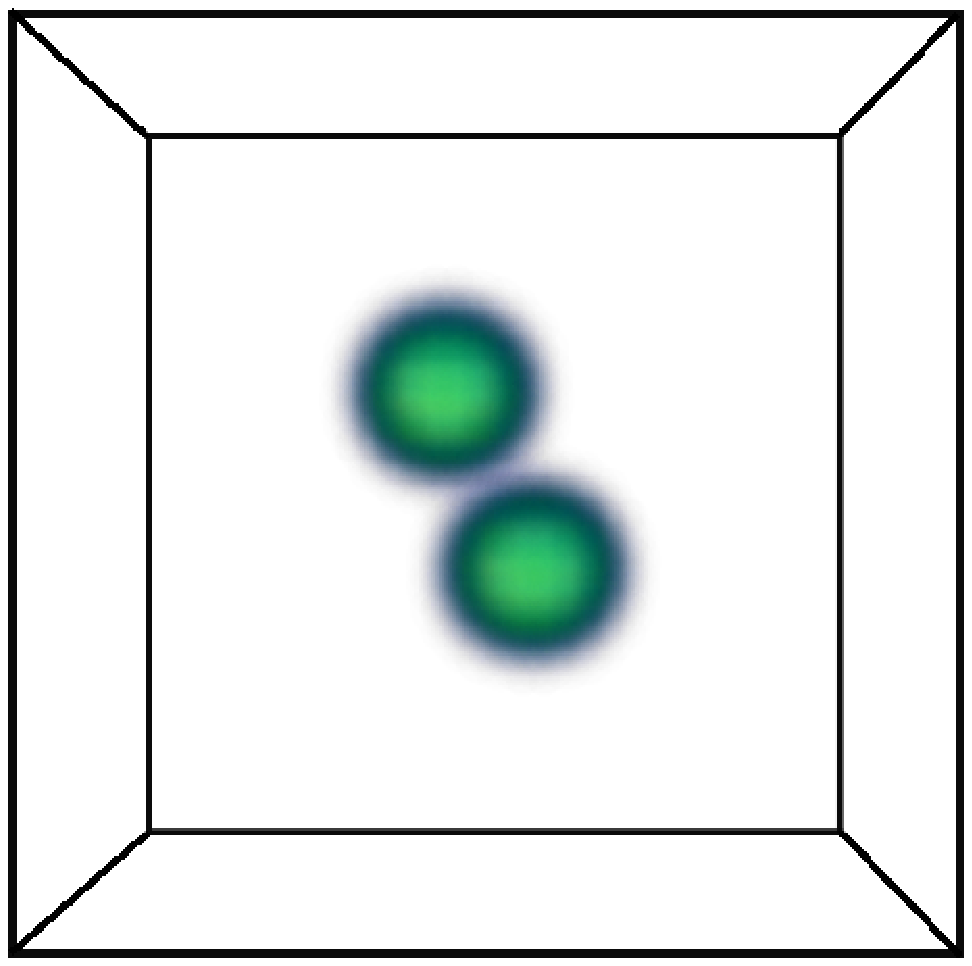}
\includegraphics[width=2.cm]{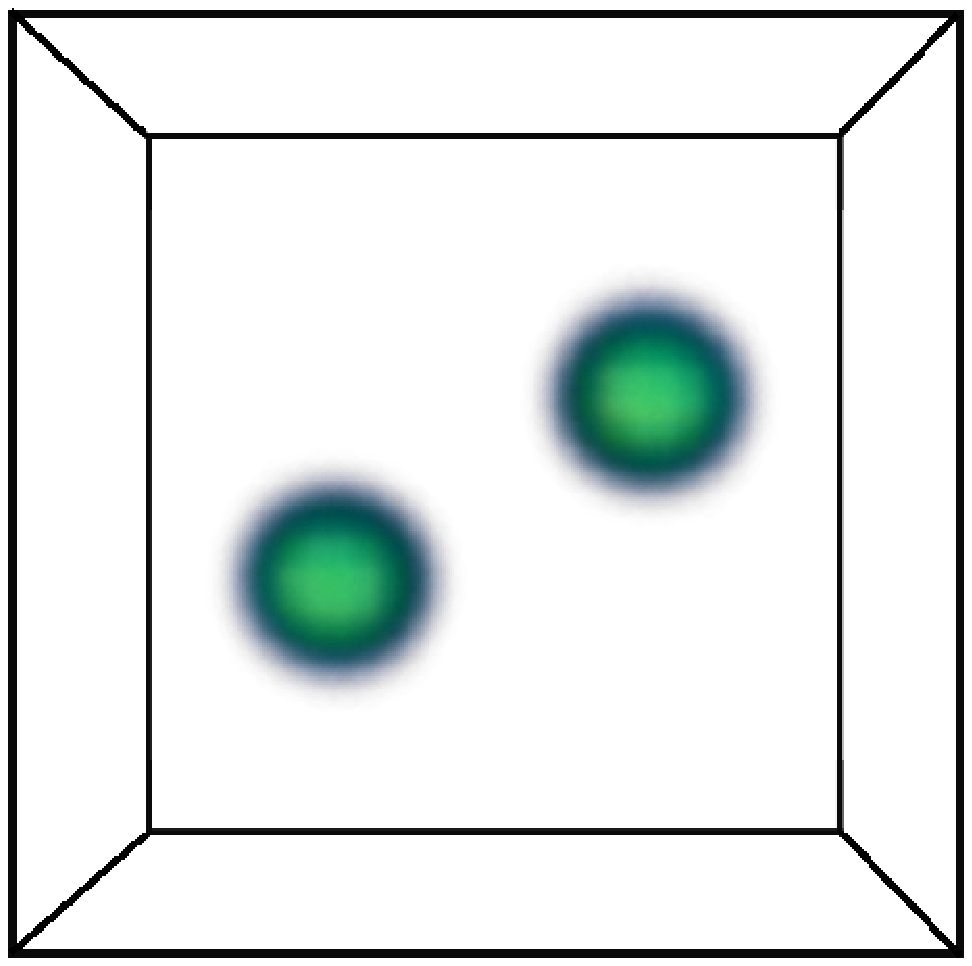}\vspace{1.5mm} \\
\caption{\label{fig4} (Color online) Two-nucleus collisions with 
different impact parameters.  The difference from Fig.~\ref{fig3}  
lies solely in the absence of Nucleus II in the initial configurations 
illustrated by Fig.~\ref{fig2}.  Snapshots at 3.33$\times10^{-22}$~s 
(left), 6.66$\times10^{-22}$~s (middle), 13.3$\times10^{-22}$~s (right) are 
shown for $|{\bm b}|$ = 1,3,5~fm.}
\end{figure}

For comparison, we consider two-nucleus collisions by starting with 
the same initial configurations as in Fig.~\ref{fig2} except for the absence 
of Nucleus II. 
The corresponding incident energy is $2.20$~MeV per nucleon in the center-of-mass frame, and the fusion product is $^{112}$Te.
 We can observe from Fig.\ \ref{fig4} that a similar but shorter structure occurs in the case of two-nucleus collisions.  
The degree of deformation from a spherical case is significantly larger in 
the case of ternary collisions (see Table \ref{table1}).
We have thus quantitatively confirmed that ternary collisions are more efficient than binary collisions in producing thin-long structures.

Note that fusion is not only one of the most efficient ways of 
producing heavy elements in stars, but also the main method for producing superheavy nuclei in the laboratory.
The present TDDFT calculations suggest that fusion reactions of two or three identical $^{56}$Fe, which lead to the synthesis of chemical elements 
heavier than iron, can be easier to occur with the help of rotations.

\begin{table}[b] 
\caption{\label{table1}
Length of the thin-long products by ternary collision at 26.6$\times10^{-22}$~s and by binary collision at 13.3$\times10^{-22}$~s
that are shown in Figs.~\ref{fig3} and \ref{fig4}.  The diameter 
for a minimum sphere that contains each product is calculated and 
then averaged over all the cases in which the product is fused.  The resultant 
length, denoted by ${\mathcal R}$, is divided by $A^{1/3}$.  For 
comparison, a typical diameter for a spherical nucleus is also given.}
\begin{tabular}{|c|c|c|c|}
\hline
\multicolumn{1}{|c|}{ } & \multicolumn{1}{c|}{$n$=3} & \multicolumn{1}{c|}{$n$=2} & \multicolumn{1}{c|}{Spherical case} \\ \hline 
${\mathcal R}/A^{1/3}$  &~   4.98~fm ~&~ 3.63~fm  ~& $\sim$2.40~fm \\  \hline  
\end{tabular}
\end{table}

\begin{figure}
(a) Model
\\
\includegraphics[width=8cm]{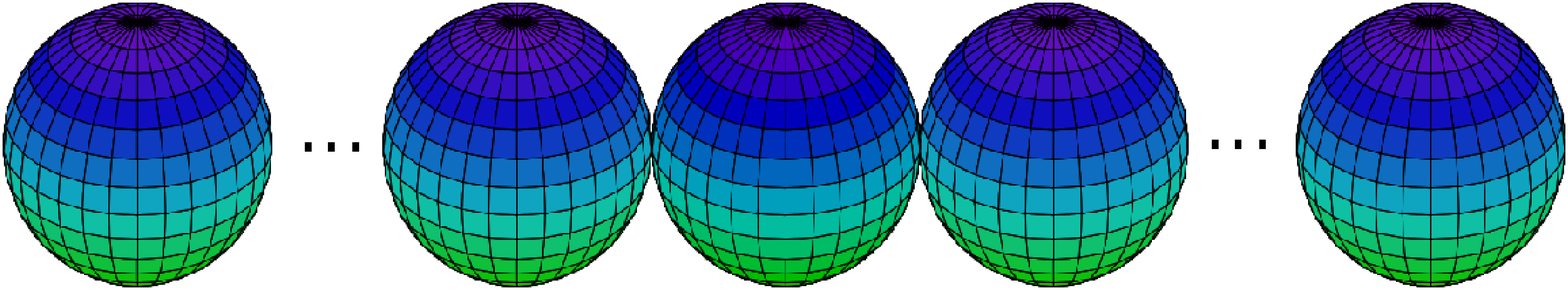} \\ 
(b)  Optimal solution
\\
\includegraphics[width=8cm]{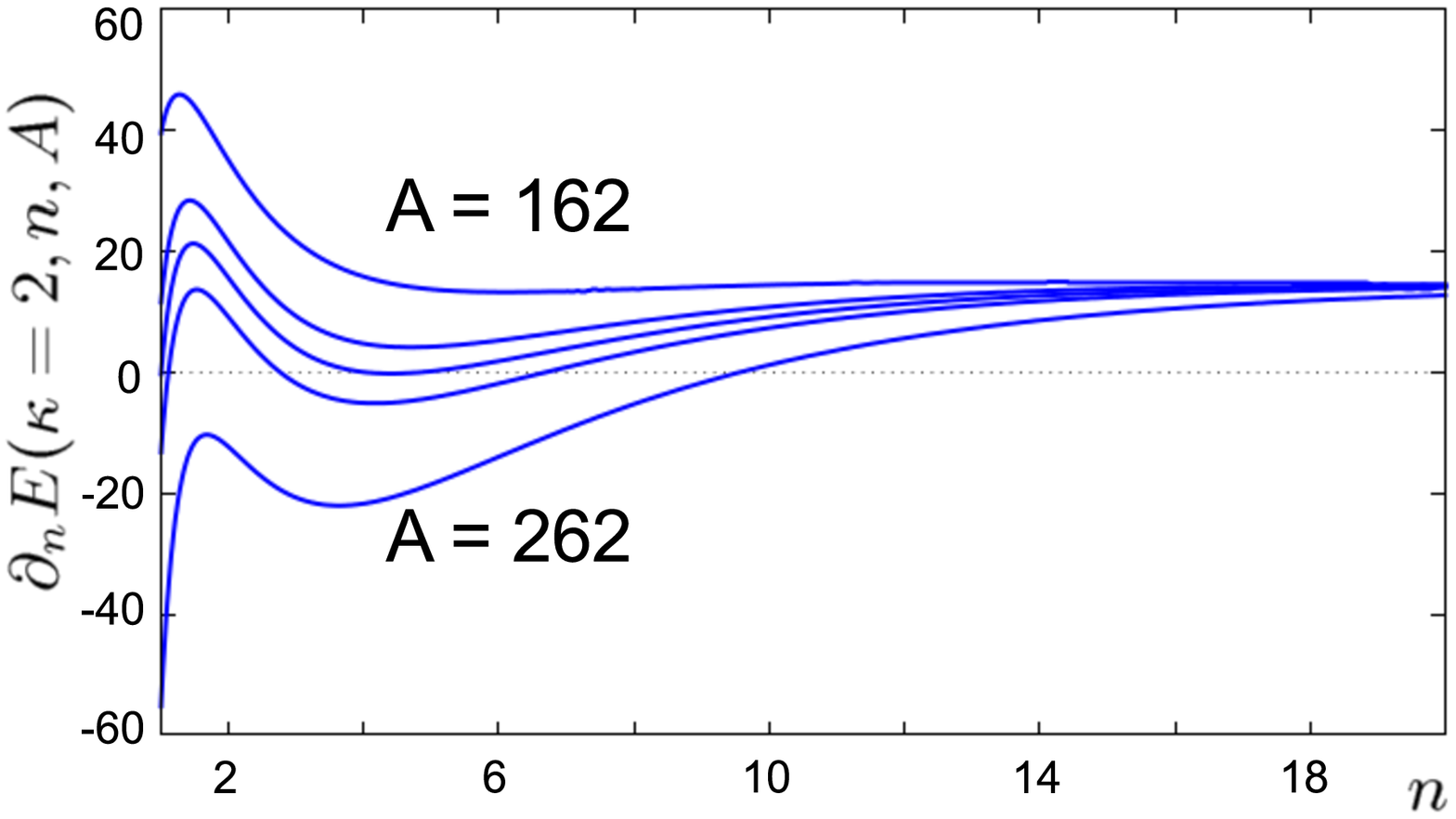} \\
\caption{\label{fig2-1} (Color online) An incompressible liquid-drop model 
for multi-nucleus collisions. 
(a) Linearly connected $n$ spherical nuclei with the total and proton numbers $(A,Z)$.
(b) $\partial_n E(\kappa=2,n,A)$ is depicted as a function of $n$ ($A =$ 162, 199, 212, 225, 262).
A state with $n = 1$ loses its global stability at $A = 225$, which 
is a little bit larger than $A_{0}(\kappa=2)=212$, and, for $A>225$,
a state with $n > 2$ is stabilized. }
\end{figure}

The efficiency of multi-nucleus collisions in the production of 
thin-long heavy nuclei can be essentially understood by the competition between the 
surface tension and the Coulomb repulsion.
Let us describe this competition within the framework of an incompressible 
liquid-drop model, which can characterize a specific geometry of the early stage 
of multi-nucleus collisions (see Fig.\ \ref{fig2-1}~(a)). In this model we 
restrict ourselves to the situation in which the total volume of
the system is fixed to $4 \pi r_0^3 A/3$ with $r_0 = 1.20$~fm and the 
number of nucleons $A$. 

Let us now consider a string of $n$ identical nuclei of radius 
$R = r_0 (A/n)^{1/3}$, which we regard as a precursor of a thin-long 
heavy nucleus as long as $n>1$ (Fig.\ \ref{fig2-1}~(a)).  
Then, we obtain the length $R_z$ of this string as
\begin{equation}
R_z = 2nR = 2r_0 n^{2/3} A^{1/3}.
\end{equation}
The liquid-drop model allows one to quantify the competition between the 
surface tension and the Coulomb repulsion, which is essential to the formation of 
thin-long heavy nuclei.  
The surface and Coulomb energies of the string can be written as
\begin{equation}
E_{\rm Surf} = a_{s} \frac{4n \pi R^2}{4 \pi r_0^2}  = a_{s} ~ n^{1/3} A^{2/3}
\end{equation}
and
\begin{eqnarray}
E_{\rm Coul} && =\sum_{i>j}  a_{c} \frac{(A/ \kappa n)^2}{r_{ij}}  + \frac{3}{5} n a_c \frac{(A/\kappa n)^2}{R}   \nonumber \\
 && = \frac{2 a_{c}}{\kappa^2 r_0}  \left[\frac{4}{5} n^{-2/3}  +  (2^{-n}-1)n^{-5/3} \right] A^{5/3},
\end{eqnarray}
where $\kappa=A/Z$ with the number of protons $Z$, $r_{ij}$ is 
the distance between the centers of the $i$-th and $j$-th nuclei that constitute 
the string, and $a_{s}$ and 
$a_{c}$ are set to $17.0$~MeV and $1.38$~MeV fm, respectively, in a 
manner that is consistent with empirical masses of stable nuclei. 
Accordingly, the competition is characterized by the sum
\begin{eqnarray} \label{liquid-energy}
E(\kappa,n,A) &\equiv&  E_{\rm Surf} +  E_{\rm Coul} \nonumber \\
 &=&  17.0~ n^{1/3}  A^{2/3} +\frac{2.30}{\kappa^{2}}  \left[\frac{4}{5} n^{-2/3}  \right.
           \nonumber \\
 &&           + \left. (2^{-n}-1)n^{-5/3} \right] A^{5/3}~({\rm MeV}). 
\end{eqnarray}
Although $n$ and $A$ are integers, for optimization we take $n$ and $A$ as real numbers.
The extremal condition for given $\kappa$ and $A$ can then be obtained from
\begin{eqnarray}
&&\partial_n E(\kappa,n,A) \nonumber \\
&=&\frac{850\kappa^2n^22^{n} -[(184n-575)2^n+239n+575]A}{150\kappa^2n^{8/3}
   2^n A^{-2/3}} \nonumber \\
&&  ({\rm MeV}).
\end{eqnarray}

\begin{table}[b] 
\caption{\label{table2}
The energy $E(\kappa,n,A)$ calculated from Eq.\ \eqref{liquid-energy}
as function of $n$.  We take $A=400,500$ and $\kappa = 2.5$, which are typical values expected for ternary collisions of heavy nuclei in the laboratory.}
\begin{tabular}{|c|c|c|c|}
\hline
\multicolumn{1}{|c|}{ } & \multicolumn{1}{c|}{$n$=3} & \multicolumn{1}{c|}{$n$=2} & \multicolumn{1}{c|}{$n$=1}  \\ \hline 
$E(2.5,n,400)$  &~   3.28~GeV ~&~ 3.30~GeV  ~&~ 3.32~GeV \\ 
$E(2.5,n,500)$  &~   4.38~GeV ~&~ 4.45~GeV  ~&~ 4.55~GeV \\  \hline  
\end{tabular}
\end{table}

The mass number satisfying $ \partial_n E(\kappa,n,A_{\rm crit}) = 0$ corresponds to
\begin{equation}
A_{\rm crit}(\kappa,n) = \frac{850\kappa^2n^22^{n}}{(184n-575)2^n+239n+575}, 
\end{equation}
where the denominator is positive for any $n$.
At least one set of the numbers $(\kappa,n,A)$ 
satisfying $\partial_n E(\kappa,n,A) = 0$ exists only when 
$A > A_{\rm crit}(\kappa,n)$.  Otherwise, $\partial_n E(\kappa,n,A) > 0$ holds.  
Since, for given $\kappa$, $A_{\rm crit}(\kappa,n)$ has the global minimum 
value $A_0(\kappa)=212 (\kappa/2)^2$, we can conclude that the value of 
$A_0(\kappa)$ gives a rough criterion for the occurrence of the
stability transition from the state with $n=1$ to the state with $n \ge 2$.
In fact, for $A<A_0(\kappa)$ where $\partial_n E(\kappa,n,A) > 0$ is satisfied for any $n$,
a single sphere ($n=1$) is energetically preferred, while for $A>A_0(\kappa)$,
as shown in Fig.\ \ref{fig2-1}(b), there can be more than one value of $n$ that fulfills $\partial_n E(\kappa,n,A) = 0$.

As a result of detailed analyses of the energy landscape, we find that a state with $n=1$ remains energetically optimal for mass number up to a value that is a little bit larger than 
$A_0(\kappa)$.
 Above this value, the optimal state is not preferably realized in two-nucleus collisions ($n = 2$) but in multi-nucleus collisions ($n \ge 3$). 
This implies that for superheavy synthesis, production of thin-long nuclei by ternary collisions ($n=3$) can be more efficient than the production 
by usual binary collisions leading to the states with $n=1$ 
or $n=2$ (Table~\ref{table2}).

In summary we have found a thin-long structure of heavy nuclei as 
a result of simultaneous three-nucleus collisions within the TDDFT 
approach. The validity of the calculations using the SLy6 
interaction has been confirmed by using the other interaction  (SKI3 \cite{reinhard}) in terms of the realization and rotational stabilization of thin-long heavy nuclei.
This study is expected to provide a 
motivation for designing a new accelerator and detector system for superheavy synthesis in which three-nucleus 
simultaneous collisions can take place.  In 
fact, in addition to the existing methods for superheavy 
synthesis that are based on ``binary'' fusion and multi-nucleon 
transfer reactions, ``ternary'' fusion reactions and
subsequent rotational stabilization of compound nuclei could
provide a novel method for the superheavy science.

It is also interesting to consider possible astrophysical 
implications.  For example, the three-nucleus fusion rate in stars
and supernova cores can be estimated by allowing for the average 
over the initial configurations through the Boltzmann factor and 
the plasma effects, i.e., electron screening effects and many-body 
Coulomb correlation effects between ionic nuclei \cite{ichimaru}.  
The latter effects act to reduce the Coulomb barrier between the
colliding nuclei and thus to enhance the fusion rate.  If the 
plasma is relatively dilute and hot as in the Sun, the plasma
effects can be safely ignored.  In this case, an extension of 
the usual Gamow rate to three-nucleus fusion applies, leading
to the fusion rate as a function of the plasma temperature $T$.  
Once the plasma becomes dense and thus strongly coupled as in 
supernova cores \cite{bethe}, the plasma effects should manifest 
themselves in the fusion rate through the three-particle static 
correlation function for electron-screened ions.  However, 
it is a challenging problem to accurately obtain the three-particle
correlation function as a function of $T$ and the plasma density 
$\rho$ \cite{SPPII}.  Recall that there are optimal values of the 
incident energy for three-nucleus fusion in vacuum.  In matter, 
the optimal values can be lowered by the plasma effects, 
leading to enhancement of the three-nucleus fusion rate through
the Boltzmann factor.  If the inverse of this rate 
at the values of $\rho$ and $T$ relevant for stars and supernova
cores is shorter than the corresponding evolutionary time scale, 
one can expect that thin-long heavy nuclei occur in such 
celestial objects.  

This work was supported in part by the Helmholtz 
Alliance HA216/EMMI and in part by Grants-in-Aid for Scientific
Research on Innovative Areas through No.\ 24105008 provided by MEXT.
Y.I. and K.I. acknowledge the hospitality of the Yukawa Institute for
Theoretical Physics, where this work was initiated during the
international molecule "Physics of structure and reaction of
neutron-rich nuclei and surface of neutron stars studied with
time-dependent Hartree-Fock approach".
Y. I. thanks Profs.\ T. Otsuka and J. A. Maruhn for encouragement.


\begin{thebibliography}{1}

\bibitem{clayton}
D. D. Clayton, {\em Principles of Stellar Evolution and 
Nucleosynthesis} (The University of Chicago Press, Chicago, 1983).

\bibitem{rnb7}
See, e.g., Proceedings of the Seventh International Conference on 
Radioactive Nuclear Beams, edited by C. Signorini, S. Lunardi,
R. Menagazzo, and A. Vitturi [Eur.\ Phys.\ J. Special Topics 
{\bf 150}, (2007)].

\bibitem{ogata}
K. Ogata, M. Kan, and M. Kamimura, Prog.\ Theor.\ Phys.\ {\bf 122}, 1055
(2009).

\bibitem{ichimaru}
S. Ichimaru, Rev.\ Mod.\ Phys.\ {\bf 65}, 255 (1993).

\bibitem{sawyer}
L. S. Brown and R. Sawyer, Rev.\ Mod.\ Phys.\ {\bf 69}, 411 (1997).

\bibitem{dirac}
P. A. M. Dirac, Proc.\ Cambridge Phil.\ Soc.\ {\bf 26}, 376 (1930).

\bibitem{Chabanat-Bonche}
E. Chabanat, P. Bonche, P. Haensel, J. Meyer, and R. Schaeffer, Nucl.\ Phys.\ {\bf A635}, 231 (1998); {\bf A643}, 441(E) (1998).

\bibitem{reinhard} 
P.-G. Reinhard, and H. Flocard, Nucl.\ Phys.\ {\bf A584}, 467 (1995).

\bibitem{iwata-prl} 
Y. Iwata, T. Otsuka, J. A. Maruhn, and N. Itagaki, Phys. Rev. Lett. {\bf 104}, 252501 (2010).

\bibitem{PTEP}
T. Maruyama, G. Watanabe, and S. Chiba, Prog.\ Theor.\ Exp.\ Phys.\
(2012) 2012 (1); 01A201.

\bibitem{bethe} 
H. Bethe, Rev.\ Mod.\ Phys.\ {\bf 62}, 4 (1990).

\bibitem{ichikawa11}
T. Ichikawa, J. A. Maruhn, N. Itagaki, and S. Ohkubo,  Phys.\ Rev.\ Lett.\ {\bf 107}, 112501 (2011).

\bibitem{SPPII}
S. Ichimaru, {\em Statistical Plasma Physics, Vol.\ II} (Addison-Wesley, Reading, 1994).


\end{thebibliography}
\end{document}